\documentclass[12pt,showpacs,twocolumn, preprintnumbers,aps,pre, reprint,superscriptaddress,nofootinbib]{revtex4-1}
\usepackage{graphicx}
\usepackage{subfigure}
\usepackage[labelfont={footnotesize,bf},textfont={footnotesize}]{caption}
\usepackage{color}
\usepackage{amsmath,amssymb,mathtools}
\usepackage{epstopdf}
\usepackage[linktocpage,colorlinks=true,linkcolor=blue,citecolor=blue, urlcolor = black, breaklinks=true]{hyperref}
\usepackage{url}
\usepackage{verbatim}
\usepackage{breakcites}
\usepackage[version=3]{mhchem}

\newcommand{\be}{\begin{equation}}
\newcommand{\ee}{\end{equation}}

\DeclareMathOperator{\Ra}{Ra}

\makeatother

\begin{document}

\preprint{}

\title{Circuit Bounds on Stochastic Transport in the Lorenz equations}

\author{Scott Weady}
\affiliation{Yale University, New Haven, USA}

\author{Sahil Agarwal}
\affiliation{Yale University, New Haven, USA}

\author{Larry Wilen}
\affiliation{Yale University, New Haven, USA}

\author{J. S. Wettlaufer}
\affiliation{Yale University, New Haven, USA}
%\affiliation{Program in Applied Mathematics, Yale University, New Haven, USA}
\affiliation{Mathematical Institute, University of Oxford, Oxford, UK}
\affiliation{Nordita, Royal Institute of Technology and Stockholm University, SE-10691 Stockholm, Sweden}
\email[]{john.wettlaufer@yale.edu}

\date{\today}

\begin{abstract}

In turbulent Rayleigh-B\'enard convection one seeks the relationship between the heat transport, captured by the Nusselt number, and the temperature drop across the convecting layer, captured by Rayleigh number.  In experiments, one measures the Nusselt number for a given Rayleigh number, and the question of how close that value is to the maximal transport is a key prediction of variational fluid mechanics in the form of an upper bound. The Lorenz equations have traditionally been studied as a simplified model of turbulent Rayleigh-B\'enard  convection, and hence it is natural to investigate their upper bounds, which has previously been done numerically and analytically, but they are not as easily accessible in an experimental context.  Here we describe a specially built circuit that is the experimental analogue of the Lorenz equations and compare its output to the recently determined upper bounds of the stochastic Lorenz equations \cite{AWSUB:2016}.   The circuit is substantially more efficient than computational solutions, and hence we can more easily examine the system.  Because of offsets that appear naturally in the circuit, we are motivated to study unique bifurcation phenomena that arise as a result.  Namely, for a given Rayleigh number, we find a reentrant behavior of the transport on noise amplitude and this varies with Rayleigh number passing from the homoclinic to the Hopf bifurcation.

\end{abstract}


\maketitle

\section{Introduction}

The Lorenz equations are an archetype for key aspects of nonlinear dynamics, chaos and a range of other phenomena that manifest themselves across all fields of science, particularly in fluid flow \cite[see e.g.,][]{Ruelle, Strogatz:2014}.  Lorenz \cite{Lorenz63} derived his model to describe a simplified version of Saltzman's treatment of finite amplitude convection in the atmosphere \cite{Saltzman:1962}.  The three coupled Lorenz equations, which initiated the modern field we now call chaos theory, are 
\begin{equation}
\begin{aligned}
&\dot x = \sigma(y-x),\\
&\dot y = \rho x - xz - y \quad \textrm{and} \\
&\dot z = xy-\beta z, 
\label{eq:Lorenz}
\end{aligned}
\end{equation}
where $x$ describes the intensity of convective motion, $y$ the temperature difference between ascending and descending fluid and $z$ the deviation from linearity of the vertical temperature profile.  The parameters are the Prandtl number $\sigma$, the normalized Rayleigh number, $\rho=\frac{\Ra}{~\Ra_c}$, where $\Ra_c=\frac{27\pi^4}{4}$, 
and a geometric factor $\beta$. Here we take $\sigma=10$ and $\beta=\frac{8}{3}$, the original values used by Lorenz. 

The sensitivity of solutions to small perturbations in initial conditions and/or parameter values characterize chaotic dynamics and have a wide array of implications. Chaotic behavior does not lend itself well to standard analysis, but modern computational methods provide us with vastly more powerful tools than those available to Lorenz. However, one powerful mathematical method used for example in the study of fluid flows is variational, and assesses the optimal value of a transport quantity, or a bound \cite{Howard:1972, DoeringBook, Kerswell:1998}, which we briefly discuss next. 

\subsection{Bounds on Fluid Flows}

Bounding quantities in fluid flows has important physical consequences and substantial theoretical significance.  Whereas variational principles are central when an action is well-defined and phase space volume is conserved, they pose significant challenges for dissipative nonlinear systems in which the phase space volume is not conserved and thus not Hamiltonian \cite[e.g.,][]{Wisdom}.  However, initiated by the work of Howard \cite{Howard:1963}, who used a variational approach to determine the upper bounds on heat transport in statistically stationary Rayleigh-B\'enard convection, with incompressibility as one of the constraints, the concept of mathematically bounding the behavior of a host of flow configurations has developed substantially \cite{Kerswell:1998}, as well as in other dissipative systems such as solidification \cite{WWO:2010}.  

Transport in the Lorenz system is defined as the quantity $\langle xy\rangle$ where $\langle\cdot\rangle$ denotes the infinite time average. We also note that this quantity is proportional to $\langle z\rangle$ and $\langle x^2\rangle$. Bounds on transport were first produced by \citet{Malkus:1972} and \citet{Edgar} in the 1970s. Knobloch used the theory of stochastic differential equations to analyze statistical behavior in the Lorenz system, with particular focus on the computation of long time averages, including the transport. His method can be seen as an early incarnation of the background method of Constantin and Doering \cite{DoeringBook}. Following the development of new analytical tools, interest in bounds in the Lorenz system their interpretation grown over the past two decades. Using the background method, \citet{Souza:2015} produced sharp upper bounds on the transport ${\langle xy \rangle\leq \beta(\rho-1)}$, which are saturated by the non-trivial equilibrium solutions ${(x,y,z)\equiv(x_0,y_0,z_0)=(\pm\sqrt{\beta(\rho-1)},\pm\sqrt{\beta(\rho-1)},\rho-1)}$. \citet{AWSUB:2016} extended their result to the stochastic Lorenz system, recovering the sharp bounds in the zero noise amplitude limit.

Recent numerical work, especially in the form of semi-definite programming, has provided novel methods for bounding and locating optimal trajectories, that is, trajectories that maximize some function of a system's state variables \cite{Cherny:2014}.

\citet{Tobasco} describe such an approach to this problem through the use of auxiliary functions similar to Lyapunov functions used in stability analyses. \citet{Goluskin} utilizes this method to compute example bounds on polynomials in the Lorenz system. For transport in particular (the polynomial $xy$), his results agree with the existing analytical theory. In the chaotic regime, however, we know the optimal solutions are unstable and are only attained for a very specific set of initial conditions, and in the stochastic system such solutions may never be realized. Hence, it is natural to ask about bounds on non-specious trajectories. \citet{Fantuzzi} present a semi-definite programming approach to this problem similar to that of \citet{Tobasco}, though work still needs to be done to apply their methods to systems containing unstable limit cycles and saddle point equilibria, which includes the Lorenz system.

We offer an alternative method for analyzing time-averaged behavior through the use of an analog circuit. Circuits can model a wide range of linear and nonlinear dynamical systems, and by collecting voltage data from the circuit we can perform calculations of any function of the systems state variables. In this paper, we use the circuit approach to study transport, $\langle xy\rangle$, in the stochastic Lorenz system, a choice which is motivated by its physical analogy with Rayleigh-B\`enard convection. For true convective motion, experimental measurements of transport are challenging, and the circuit provides us with a quick and easy way to perform these calculations, in fact much faster than standard numerical methods. We first introduce the stochastic Lorenz system and the corresponding bounds on transport. We then discuss the circuit implementation and offer an analytical model for the circuit system. Finally, we discuss our computations of transport in relation to the analytical upper bound theory and compare our results to the numerical solutions.

\section{The Circuit Lorenz Experiment}
\noindent
\subsection{Upper Bounds of the Stochastic Lorenz System}

The Lorenz system might be best described as a {\em motif} of atmospheric convection, which was the motivation for its derivation.  However, such physically based models can often become more realistic by adding a stochastic element to account for random fluctuations, observational error, and unresolved processes. This conceptually common idea has become particularly popular in climate modeling and weather prediction \cite[e.g.,][]{Arnold:2013aa, Majda:2017}.  Here, we follow this approach in  the Lorenz system by adding a stochastic term with a constant coefficient \cite{AWSUB:2016} viz., 
\begin{equation}
\begin{aligned}
&\dot x = \sigma(y-x) + A \xi_x,\\
&\dot y = \rho x - xz - y+A \xi_y \quad \textrm{and} \\
&\dot z = xy - \beta z + A \xi_z, 
\label{eq:StochasticLorenz}
\end{aligned}
\end{equation}
where the $\xi_i$ are Gaussian white noise processes, $A$ is the noise amplitude and $\sigma, \beta$, and $\rho$ are as in Equations (\ref{eq:Lorenz}). 
%Throughout this paper we assume ${A_x=A_y=A_z=A}$.
%
%\subsection{Stochastic Upper Bounds}
%Our study of the circuit model is motivated by the analytical stochastic upper bound theory. 
The circuit described below in \S \ref{sec:Circuit} allows us to experimentally test and analyze stochastic bounds of the transport in the Lorenz system subject to forced and intrinsic noise. In the infinite time limit, the stochastic upper bounds of \citet{AWSUB:2016} are given by
\begin{align}\label{eq:SUBs}
\langle x y \rangle_T \leq \beta(\rho-1) + \frac{A^2}{\rho-1}\Big(1 + \frac{1}{2\sigma}\Big).  
\end{align}
For $A=0$ these reduce to the upper bounds of \citet{Souza:2015}.  However, unlike the deterministic case, the fixed point solutions do not exist so that the optimum is never truly attained.  We note that these bounds tend to infinity as $\rho\rightarrow1$, though \citet{FantuzziGFD} improved this bound in the low Rayleigh number regime. 

\subsection{The Lorenz Electrical Circuit\label{sec:Circuit}}

\begin{figure}[h]
\includegraphics[width=\columnwidth]{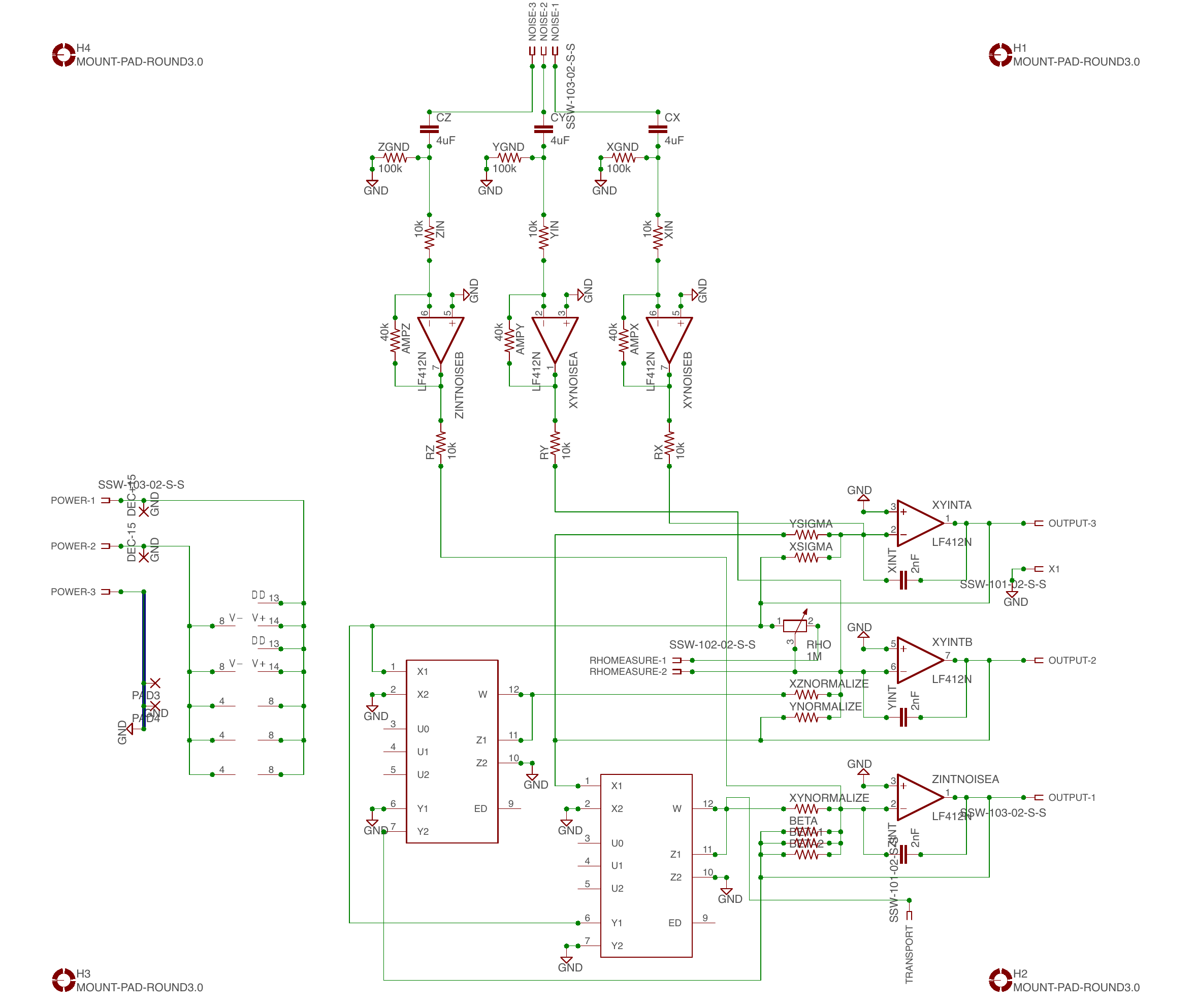}
\centering
\caption{Schematic of the stochastic Lorenz circuit.}
\label{fig:Circuit}
\end{figure}

Following the implementation described by Horowitz \cite{Horowitz}, the Lorenz system is modeled in an analog circuit through a series of op-amp integrators and voltage multipliers (Fig. \ref{fig:Circuit}). Mathematically, this implementation essentially solves Equations (\ref{eq:Lorenz}) by continuously integrating both sides and returning the output $x,y,z$ back into the circuit. Adding a noise element to the integrators allows us to adapt this circuit to the stochastic Lorenz systems. To generate noise we use Teensy 3.5 microprocessors. These boards possess hardware random number generators that provide a higher quality of randomness compared to those more commonly found on microprocessors and computers. They have 12-bit resolution digital to analog converters (DAC) allowing us to output a voltage between 0V and 3.3V at $2^{12} = 4096$ discrete values. This gives us better spectral characteristics compared to pulse-width modulation which outputs either 0V or 3.3V with a duty cycle that corresponds to the analog level. To achieve Gaussian random noise we sum 8 random integers, chosen in a limited range corresponding to the noise amplitude. The number is then centered about the middle voltage corresponding to the integer 2048 and outputed through the DAC channel. Following this process the signal is AC coupled to ensure the voltage is symmetric about 0V, and further amplification is achieved through an op-amp. This method allows us to easily control the noise processes and amplitudes directly from the computer, and thus to automate many components of the experiment.

To collect voltage data from the circuit, we use an Arduino Due microprocessor with 12-bit analog read resolution along with several voltage dividers and amplifiers to put the voltages in the Arduino's range of 0-3.3V. As in the noise generation, when processed the voltages appear as an integer between 0 and 4095, corresponding to a voltage between 0-3.3V. From this data we can convert back to the original voltage using measurements of the amplifiers and voltage dividers and scaling by 10, the normalization factor of the circuit.

\begin{figure*}[t]
\centering
(a)\includegraphics[trim={0cm 7cm 0cm 5cm},width=0.98\columnwidth]{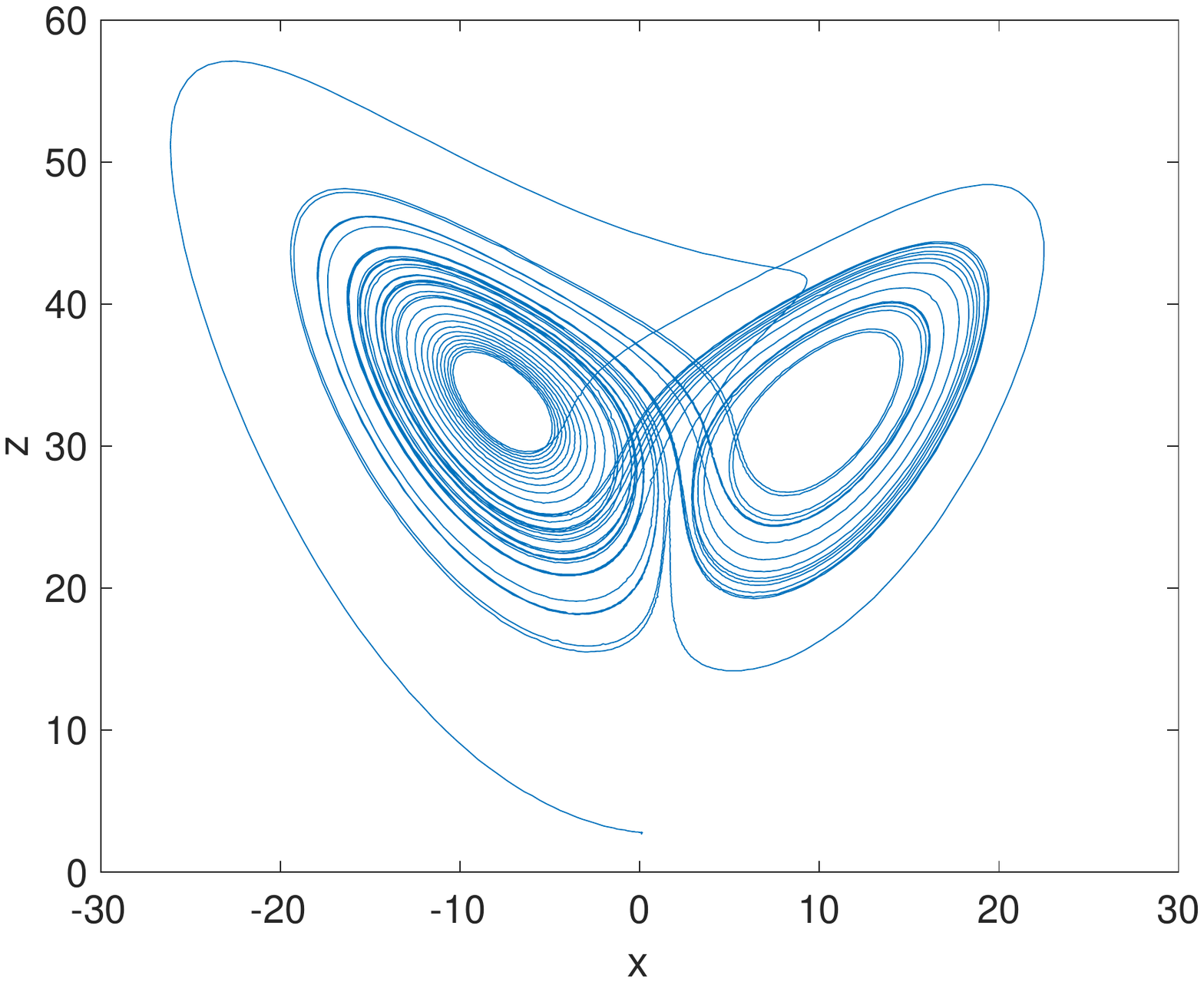}
\centering
(b)\includegraphics[trim={0cm 7cm 0cm 5cm},width=0.98\columnwidth]{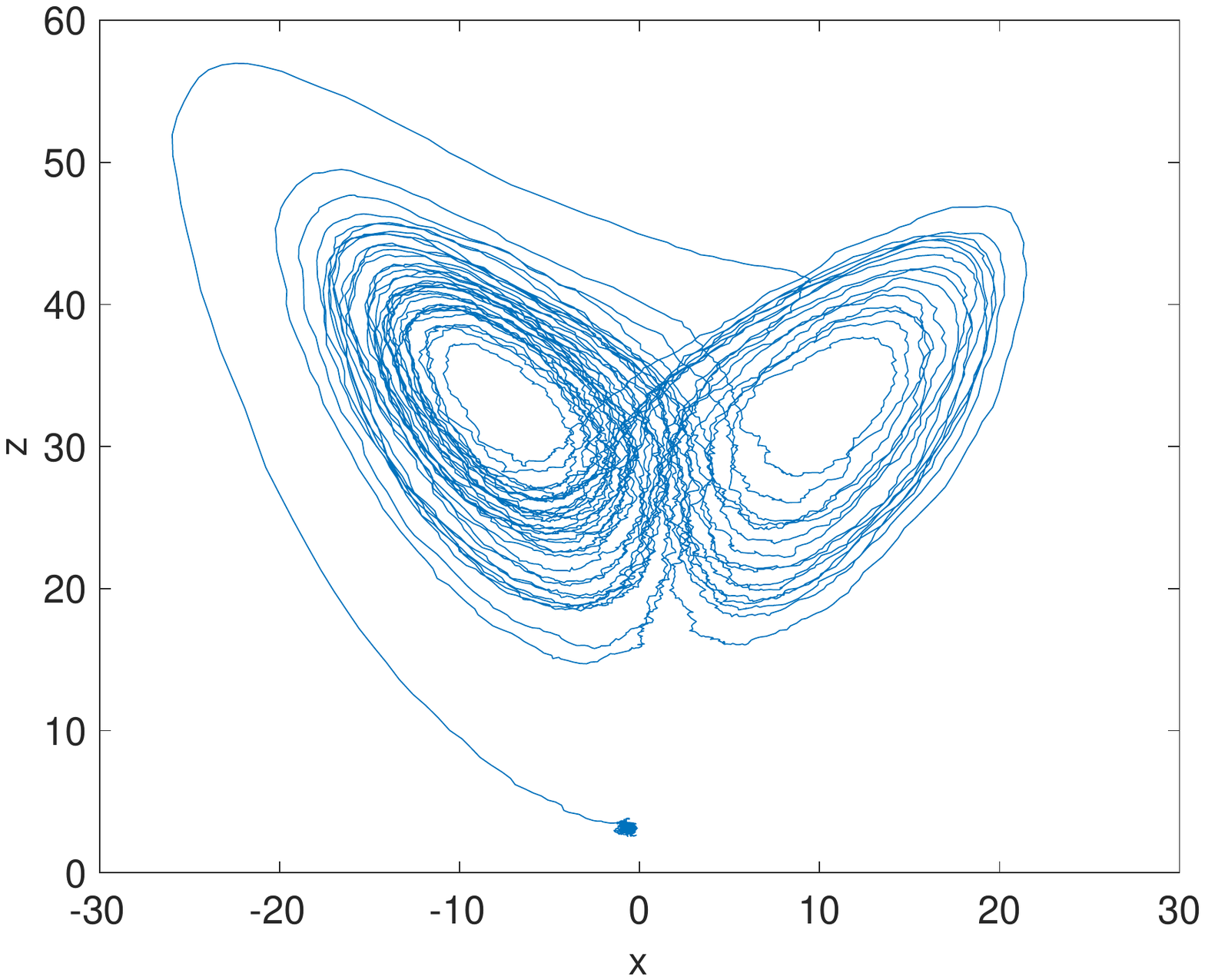}
%\includegraphics[trim={0cm 7cm 0cm 5cm},scale=0.42]{AttractorA4.pdf}
%\centering
\caption{Circuit generated stochastic Lorenz attractor in the $x-z$ plane for (a) $A=0$ and, (b) $A=4$.}
\label{fig:CircuitLorenz}
\end{figure*}

The rate of integration is determined by the three capacitors, ideally equal in value. This allows us to adjust the sampling rate depending on the application. For measuring transport, we can run the circuit at a very high speed and sample as fast as possible, approximately every 100$\mu$s with the Arduino Due. Figure \ref{fig:CircuitLorenz} shows samples of the circuit-generated attractor for noise amplitudes ${A=0}$ and ${A=4}$. To achieve the initial condition we ground the $y$ integrator while the circuit is powered up and then close this connection to start solving the system. We note that the noise amplitudes are chosen in reference to a baseline voltage and do not numerically correspond to the same amplitude in Equations (\ref{eq:StochasticLorenz}).

A key feature of the circuit is the ability to sample $xy$ directly from the evolution equation for $z$. Not only does this provide much faster convergence of $\langle xy\rangle_T$, but keeping a running average avoids the need to store large arrays of data or perform extra arithmetic operations.

\section{The Offset Lorenz System}

Any analog circuit is subject to non-idealities due to input and output offsets in op-amps and multipliers and inaccurate measurements of circuit elements, resulting in non-ideal gain factors. There are also contributions to the noise from all of these components. Since the Lorenz system is symmetric under the transformation $(x,y,z)\mapsto(-x,-y,z)$, these non-idealities introduce asymmetry into the solutions of Equations (\ref{eq:Lorenz}), which poses an issue for our model. To account for this asymmetry we propose a slightly modified set of equations for the circuit model and analyze their properties.

It is easily demonstrated that the largest contribution is from output offsets in the multipliers, and we therefore simplify the analysis by neglecting all other sources of error or noise. We formally define the offset Lorenz system by
\begin{equation}\label{eq:offset}
\begin{aligned}
\dot x &= \sigma(y - x),\\
\dot y &= \rho x - xz - y + \varepsilon_y \quad \textrm{and}\\
\dot z &= xy - \beta z + \varepsilon_z,
\end{aligned}
\end{equation}
where $\varepsilon_y$ and $\varepsilon_z$ are small constant offsets resulting from the product terms.

\subsection{Fixed Points \& Stability}

It is straightforward to see that the equilibria are given by
\begin{equation}
x^* = y^* = \pm\sqrt{\beta z^* - \varepsilon_z},\\
\end{equation}
where $z^*$ is a solution to the cubic equation
\begin{equation}\label{eq:zpoly}
\begin{aligned}
0 &= \beta z^3 - [2\beta(\rho - 1) + \varepsilon_z]z^2+ [2(\rho - 1) + \beta(\rho - 1)^2]z  \\ & \quad - [(\rho - 1)^2\varepsilon_z + \varepsilon_y^2].
\end{aligned}
\end{equation}

There is always one real solution to Equation (\ref{eq:zpoly}) corresponding to the trivial equilibrium, and for $\rho \gg |\varepsilon|$ a pair of roots exists corresponding to the non-trivial equilibria. We solve Equation (\ref{eq:zpoly}) numerically in order to determine the stability of the equilibrium solutions.
% For the relevant case of $\rho>1$, we assume equilibrium solutions to Equations (\ref{eq:offset}) of the form
% \begin{equation}
% \begin{aligned}
% x^* &= x_0 + \delta_x,\\
% y^* &= y_0 + \delta_y\quad \textrm{and} \\
% z^* &= z_0 + \delta_z,
% \label{eq:pertFP}
% \end{aligned}
% \end{equation}
% where $(x_0,y_0,z_0)=(\pm\sqrt{\beta(\rho-1)},\pm\sqrt{\beta(\rho-1)},\rho-1)$ are the two non-trivial equilibria of Equations (\ref{eq:Lorenz}), and 
% $\delta_x,\delta_y,$ and $\delta_z$ are small. Substituting these fixed points into Equations (\ref{eq:offset}) \textcolor{red}{and neglecting higher order terms in $\delta$, we find, to the order of the approximations,

% \begin{equation}\label{eq:equi}
% \begin{pmatrix}
% 0 \\ -\varepsilon_y \\ -\varepsilon_z
% \end{pmatrix} 
% =
% \begin{pmatrix}
% -\sigma & \sigma & 0\\
% \rho - z_0 & - 1  & -x_0\\
% y_0 & x_0  & -\beta
% \end{pmatrix}
% \begin{pmatrix}
% \delta_x \\\delta_y \\ \delta_z
% \end{pmatrix},
% \end{equation}
% \begin{equation}\label{eq:equi}
% \begin{matrix}
% 0 & = & -\sigma\delta_x & + \sigma\delta y, & \\
% -\varepsilon_y & = & (\rho - z_0) \delta_x  & - \delta_y   & - ~x_0\delta_z,\\
% -\varepsilon_z & = & y_0\delta_x & + x_0\delta_y  & - ~ \beta\delta_z.
% \end{matrix}
% \end{equation}
% which we can easily solve given $\varepsilon_y$ and $\varepsilon_z$.}

\begin{figure}[h]
\includegraphics[trim={0cm 6.5cm 0cm 7cm},width=\columnwidth]{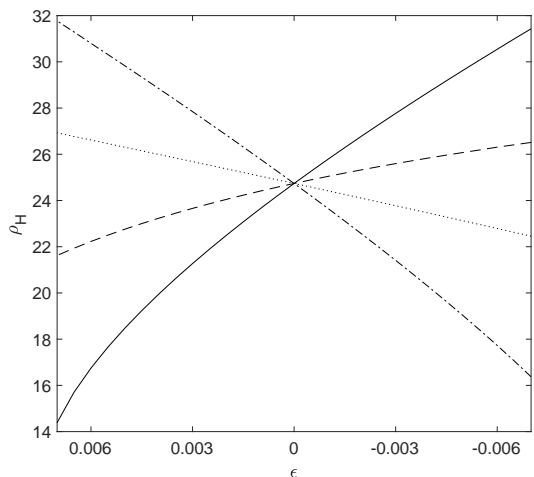}
\caption{Bifurcation value of the two non-trivial equilibria as a function of $
\varepsilon$ for $\varepsilon_y = \varepsilon_z$ (solid, $x^*_+$, and dotted, $x^*_-$, curves) and $\varepsilon_y = -\varepsilon_z$ (dashed, $x^*_+$, and dash-dotted, $x^*_-$). The $x$-axis is rescaled to $\varepsilon / 1000$ in correspondence with the equivalent voltage values. The $y$ intercept occurs at $(\varepsilon,\rho_H)=(0,24.74)$ as expected. }
\label{fig:bif}
\end{figure}

\subsubsection{Stability of the Equilibrium Solutions}

The Jacobian for our offset system is the same as the Jacobian of Equation (\ref{eq:Lorenz}):
\begin{equation}\label{eq:Jacobian}
\resizebox{0.6\linewidth}{!}{
$J(x,y,z) = \begin{pmatrix}
-\sigma & \sigma & 0\\
\rho - z & -1 & -x\\
y & x & -\beta
\end{pmatrix}$}.
\end{equation}

% \textcolor{red}{We solve the linear system (\ref{eq:equi}) numerically and substitute the solutions into the Jacobian to determine stability of the fixed points (\ref{eq:pertFP}).} To simplify the analysis, let $|\varepsilon_y| = |\varepsilon_z|$.
% Evaluating the Jacobian at the trivial offset equilibria yields
% \begin{equation}
% \resizebox{0.6\linewidth}{!}{
% $J(x^*,y^*,z^*) = \begin{pmatrix}
% -\sigma & \sigma & 0\\
% \rho - \delta_z & -1 & -\delta_x\\
% \delta_y & \delta_x & -\beta
% \end{pmatrix}.$
% }
% \end{equation}

To simplify the analysis we assume $|\varepsilon_y| = |\varepsilon_z| = \varepsilon$. We find there is an imperfect pitchfork bifurcation with respect to $\rho$, the critical value of which varies with $\varepsilon_y$ and $\varepsilon_z$. The imperfection is a natural result of symmetry breaking due to the offset terms.

At the non-trivial equilibria (assuming $\rho\gg|\varepsilon|$ so that these solutions exist) we find that, as in the standard Lorenz system, there is a Hopf bifurcation with respect to $\rho$ that now depends on $\varepsilon$ and the relationship between the signs of $\varepsilon_y$ and $\varepsilon_z$. Figure \ref{fig:bif} shows the bifurcation value $\rho_H$ of $x^*_+$ and $x^*_-$ for $\varepsilon_y = \varepsilon_z$ and $\varepsilon_y = -\varepsilon_z$. Unlike the standard Lorenz system, the bifurcation value of each equilbrium point is different, and the maximum bifurcation value of the pair of equilibria for the given relationships between $\varepsilon_y$ and $\varepsilon_z$ is greater than $24.74$, the critical value in the Lorenz system. Because the Hopf bifurcation denotes the appearance of chaos, this implies that any offset can delay the onset of chaos in our modified system. 

\subsubsection{Numerical Solutions of the Offset Lorenz System}

\begin{figure}[t!]
\includegraphics[trim={0cm 7cm 0cm 5cm},width=\columnwidth]{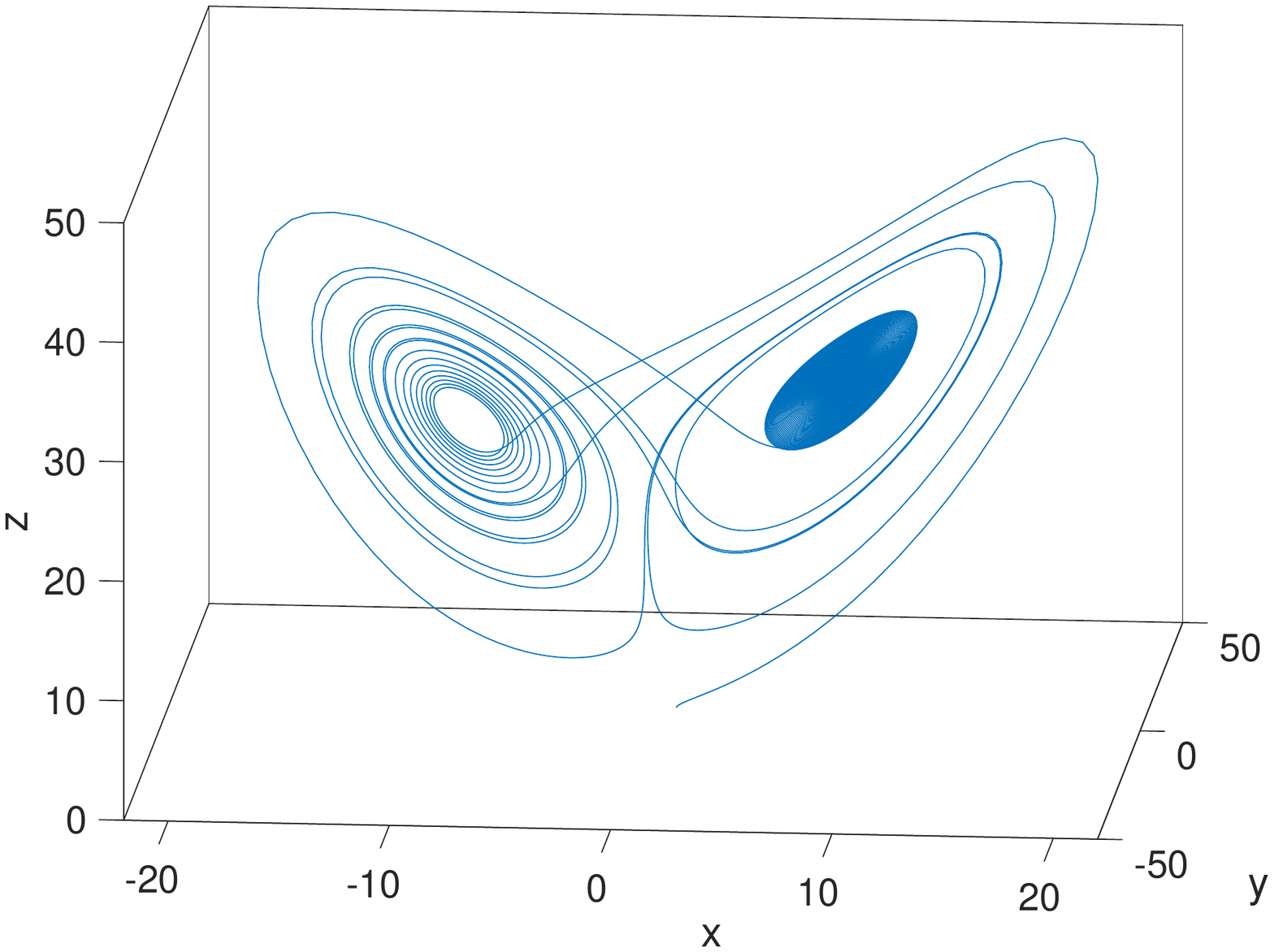}
%\centering
\caption{Plot in $(x,y,z)$ phase space of numerical simulation of the offset Lorenz system at $(\varepsilon_y/1000,\varepsilon_z/1000,\rho)=(0.005,0.005,28)$.}
\label{fig:Below}
% \end{figure}
% \begin{figure}[h]
\includegraphics[trim={0cm 7cm 0cm 5cm},width=\columnwidth]{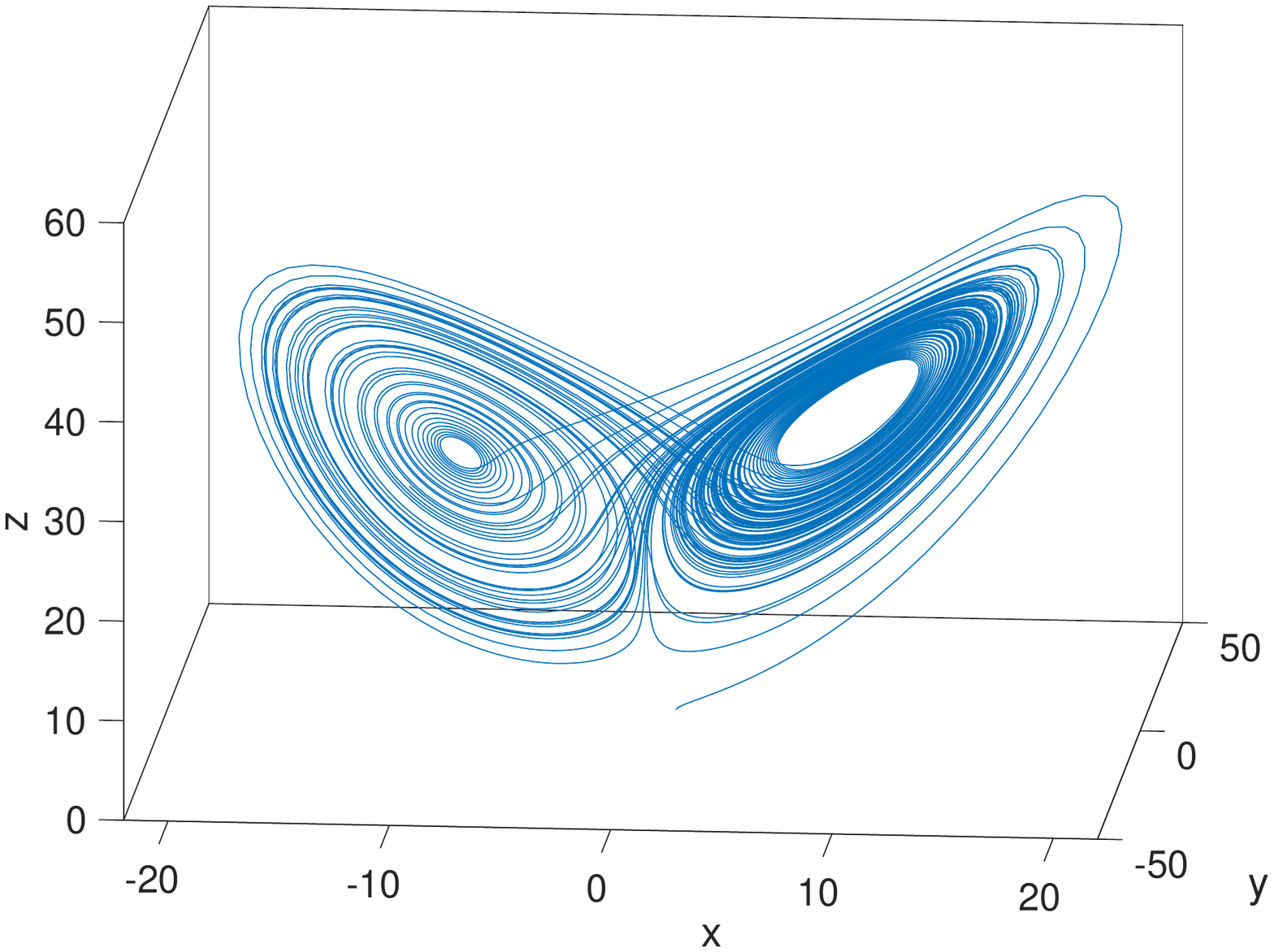}
%\centering
\caption{Plot in $(x,y,z)$ phase space of numerical simulation of the offset Lorenz system at $(\varepsilon_y/1000,\varepsilon_z/1000,\rho)=(0.005,0.005,30)$.}
\label{fig:Above}
\end{figure}

We confirm the existence of a Hopf bifurcation for the offset system using numerical solutions. For example, with $\varepsilon = 5$ and $\varepsilon_y = \varepsilon_z$, our solutions predict that the bifurcation occurs at $\rho_H = 29.83$ for $x^*_+$ and $\rho_H = 22.76$ for $x^*_-$. Figures \ref{fig:Below} and \ref{fig:Above} show numerical simulations of the offset Lorenz system slightly below and above the larger bifurcation value, respectively. We see the trajectory indeed becomes chaotic and that there is a strong preference for the positive side of the attractor, reflecting the asymmetric bifurcation values. 

\begin{figure}[hb!]
\includegraphics[trim={1cm 5cm 1cm 5cm},width=0.8\columnwidth]{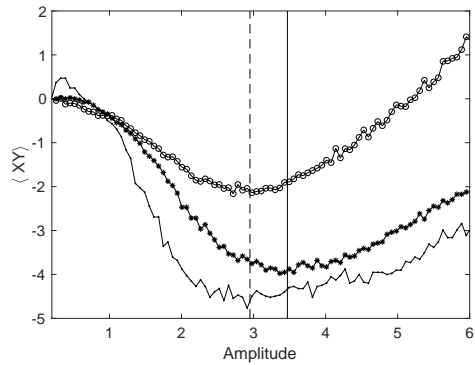}
%(a)
%\includegraphics[trim={1cm 5cm 1cm 5cm},width=0.8\columnwidth]{ampTrans_6.pdf}\\
%(b)
%\includegraphics[trim={1cm 5cm 1cm 5cm},width=0.8\columnwidth]{ampTrans_14.pdf}\\
%(c)
%\includegraphics[trim={1cm 5cm 1cm 5cm},width=0.8\columnwidth]{ampTrans_20.pdf}
\caption{Transport $\langle xy\rangle$ as a function of noise amplitude for $\rho$ = 6 (circles), $\rho$  = 14 (asterisks), and $\rho$ = 20 (dots). In each case
transport achieves a minimum value before increasing roughly linearly with noise amplitude. The location of this minimum increases until $\rho$ reaches the homoclinic bifurcation after which it decreases to zero as $\rho$ approaches the Hopf bifurcation. The solid
vertical line denotes the minimum of the $\rho$ = 14 curve and the dashed vertical line denotes the minima of the $\rho$ = 6 and $\rho$ = 20 curves. 
 \label{fig:trans_amp}}
\end{figure}

\begin{figure*}[ht]
\includegraphics[trim = {0cm 3cm 0cm 3cm}, scale=0.6]{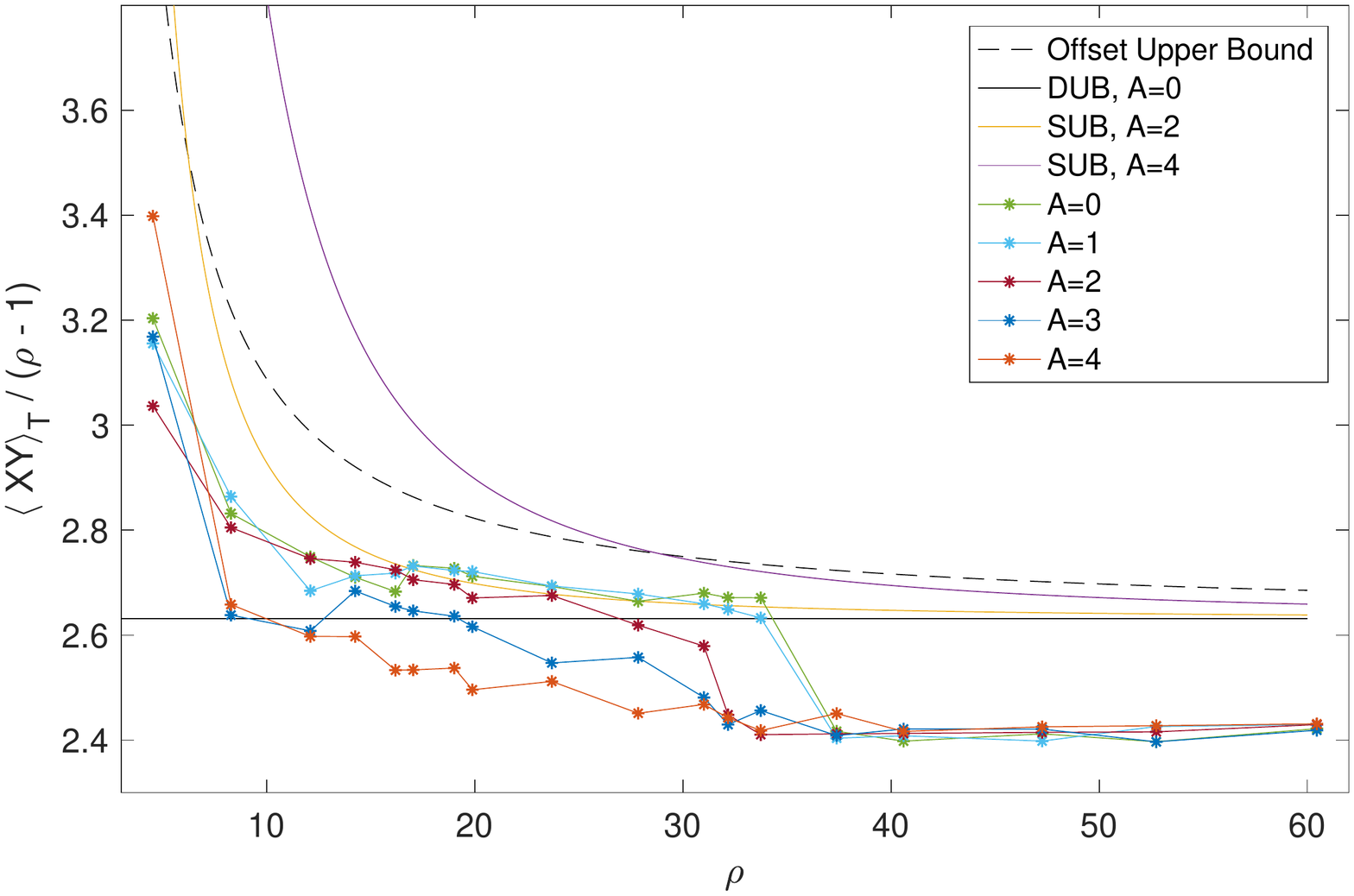}
%\centering
\caption{The scaled transport $\langle xy\rangle / (\rho - 1)$ versus $\rho$ for circuit solutions and scaled upper bounds. The transition to chaos occurs at $\rho\approx 34$. The maximum offset upper bound under the specifications of the circuit elements occurs for $\varepsilon_y = 2.5$ and $\varepsilon_z = -2.5$, which is shown as the dashed line.\label{fig:UB}}
\end{figure*}

It is important to emphasize that our discussion of the offset Lorenz system has implications for the circuit model. Most notably, it implies the transition to chaos can occur at a larger value of $\rho$ than in the standard Lorenz equations. The circuit does in fact demonstrate this, and the value at which the transition occurs varies  with the circuit elements chosen. A detailed analysis of the circuit, taking into account the gain factors in the multiplier and op amps, shows that the offset values calculated in the stability analysis must be divided by 1000 to correspond to the appropriate voltage offsets in the circuit. Hence, in Figure \ref{fig:bif}, the bifurcation values are plotted in terms of the actual voltage offsets of the multipliers. The specifications for the multiplier output offsets are $\pm$25mV, and the bifurcation values typically observed for the circuit are consistent with these specifications.

% \begin{figure*}[t]
% (a)\includegraphics[trim={1cm 5cm 1cm 5cm},scale = 0.4]{numerical_convergence_33.pdf}
% (b)\includegraphics[trim={1cm 5cm 1cm 5cm},scale = 0.4]{circuit_convergence_33.pdf}
% \centering
% \caption{}
% \end{figure*}

\section{Results \& Discussion} 

\subsection{Stochastic Upper Bounds}\par\medskip

The upper bounds in \cite{AWSUB:2016} are no longer valid in the offset system, but using a similar argument we can produce new bounds which, in the deterministic case, are given by
\begin{equation}
\langle xy\rangle \leq \beta\frac{\delta^*(\rho - \frac{\varepsilon_z}{\beta} - 1)^2}{\delta^*(\rho - \frac{\varepsilon_z}{\beta}) - 1} + \frac{\delta^*\varepsilon_y^2}{4(\delta^*(\rho - \frac{\varepsilon_z}{\beta}) - 1)(1 - \delta^*)},
\end{equation}
where $\delta^*$ is the minimizer of the function on the right hand side, which approaches 1 as $\varepsilon_y$ and $\varepsilon_z$ go to zero, yielding the sharp non-offset upper bound. The full derivation can be found in Appendix \ref{ap:upperbounds}.

Using the circuit, we reproduce the results of \citet{AWSUB:2016} for the stochastic upper bounds.  Figure \ref{fig:UB} shows the scaled average transport $\langle xy \rangle / (\rho - 1)$ for 18 values of $\rho$ and $5$ noise amplitudes, including $A=0$. The results remain close to the analytical upper bounds until the transition to chaos at $\rho\approx 34$, which is consistent with the analytical work to which we compare our approach.

At subcritical Rayleigh numbers for small noise amplitude the transport achieves a minimum value,  then increases monotonically with noise amplitude as found by \citet{AWSUB:2016}.  The amplitude at which  this minimum occurs increases until $\rho$ reaches the homoclinic bifurcation after which it decreases to zero as $\rho$ approaches the Hopf bifurcation. This behavior is shown in detail in Figure \ref{fig:trans_amp}, which we discuss further in the following section. Beyond the transition to chaos however, the increase or decrease of transport with noise amplitude at a given Rayleigh number depends largely on the sampling frequency, number of realizations, and even further by sampling resolution. Unlike numerical solutions, the circuit solves the Lorenz equations in real time, so when we sample $xy$ from the circuit we are in fact sampling from the full attractor, not just at set discrete time steps. In conjunction with the high sampling rate we might hope to see ergodic characteristics in our computations, however our sampling only shows consistent values of $\langle xy\rangle$ at very low noise amplitudes, and this is most likely a result of the resolution rather than the circuit's actual reflection of the chaotic set. Hence, we find that the behavior of the circuit in the chaotic regime is no different from the numerics, both of which show that noise is indistinguishable from chaos.

% \begin{figure}[h]
% \includegraphics[trim={1cm 6.6cm 1cm 6.6cm},width=\columnwidth]{med_transamp.pdf}
% \centering
% \caption{Transport as a function of noise amplitude for $\rho\approx 21$, produced by the circuit. The $x$-axis is the bit signal communicated by the Arduino, which roughly corresponds to $24A$. Transport achieves a minimum value before increasing linearly with amplitude. \label{fig:trans_amp}}
% \end{figure}

\subsection{Noise, Unstable Periodic Orbits, and Dissipation}

As we increase noise amplitude the relationship between $\rho$ and the transport becomes increasingly smooth, eliminating the kink at the transition to chaos. This behavior suggests the existence of a critical noise amplitude, namely, the minimum amplitude for which this transition is smooth. Taking advantage of the circuit's computational speed we can quantify this in more detail. Figure \ref{fig:trans_amp} shows circuit transport as a function of noise amplitude for $\rho = 6$, $\rho = 14$, and $\rho = 20$. As mentioned above, we find that, on average, as the noise amplitude increases, the transport achieves a minimum value before increasing linearly with amplitude. This critical noise amplitude may be interpreted as the point where noise is effectively simulating chaos, forcing the system to oscillate aperiodically about the two non-trivial fixed points of the deterministic system. The corresponding critical amplitude varies with $\rho$, increasing until the homoclinic bifurcation is reached at $\rho\approx 13.96$, after which it decreases to zero as $\rho$ approaches the Hopf bifurcation. It is for this noise amplitude that the smooth transition to chaos noted above occurs. Mathematically, a homoclinic bifurcation implies the generation of a dense set of unstable periodic orbits embedded in the attracting set, and the decrease in the critical noise amplitude corresponds to the coupling of noise with these orbits. 

\vspace{-0.5 cm}
\section{Conclusion}

Circuits modeling dynamical systems have largely been used for pedagogical purposes. Much less work, however, has utilized these circuits for actual computation and analysis. Our study of the Lorenz circuit demonstrates the value of using analog circuits for both qualitative and statistical analyses of dynamical systems. Whereas in numerical simulations one is forced to run computations for many, often very small, time steps in order to capture the useful statistics, the circuit allows one to sample directly from the attractor's ``distribution,'' meaning we need only collect data until the solution no longer fluctuates beyond a given error threshold. For the Lorenz system in the chaotic regime, we found transport converged in the circuit in approximately 4 seconds. The same calculation can take up to 5 minutes numerically to ensure accurate statistics, making the circuit approximately 100x faster. Reliably reproducing a plot such as Figure \ref{fig:trans_amp} would thus require many hours of computation numerically and about 10 minutes with the circuit. It is also worth noting that the circuit is unaffected by increasing the degrees of freedom whereas the computational cost of numerical simulation scales approximately with the order of the method.

In this paper we only discussed the influence of Gaussian white noise, though we have developed code that generates correlated noise signals, particularly pink and brown noise. Generating these signals in numerical solutions significantly increases the run time, whereas the circuit's speed is unaffected. However we found that transport behavior is more or less equivalent under the influence of colored noise, though the visible dynamics may vary. 

Ultimately an analog circuit forms a dynamical system that can be modeled by a set of differential equations. Solving the inverse problem -- constructing a circuit to fit a given set of equations -- reveals the possibility of using circuits to analyze a variety of dynamical systems of both mathematical and physical interest. Basic models have been constructed for well-studied low-dimensional systems such as the van der Pol oscillator and the R\"ossler system, but with the assistance of machine printed circuits we may be able to model much more complex dynamics extending from neuronal networks to geophysical flows. 

% Indeed, because numerical methods suffer from a wide range of instabilities, large memory requirements, and high computational costs, analog circuits eliminate these issues, offering a new and efficient way to study quantitative and qualitative behavior and experimentally test analytical theories and closure schemes.  
\vspace{-0.5 cm}

\section{Acknowledgements}

All of the authors thank Yale University for support. 
SA and JSW acknowledge NASA Grant NNH13ZDA001N-CRYO for support. JSW acknowledges Swedish Research Council Grant 638-2013- 9243 and a Royal Society Wolfson Research Merit Award for support.
\vspace{-0.5 cm}

\appendix
\section{Offset Upper Bounds}\label{ap:upperbounds}

We closely follow the argument given in reference \cite{Souza:2015}. It is first convenient to make the change of variables
\begin{equation}
\begin{aligned}
x &= x',\\
y &= \rho'y'\quad \text{and}\\
z &= \rho'z' + \frac{\varepsilon_z}{\beta},
\end{aligned}
\end{equation}
where $\rho' = \rho - \frac{\varepsilon_z}{\beta}$. Equation (\ref{eq:Lorenz}) then becomes
\begin{align}
\dot x' &= -\sigma x' + \sigma\rho' y',\\
\dot y' &= x' - x'z' - y' + \frac{\varepsilon_y}{\rho'},\\
\dot z' &= x'y' - \beta z'.
\end{align}

Averaging time derivatives of $\frac{1}{2}x'^2$, $\frac{1}{2}(y'^2 + z'^2)$, and $-z'$ (see \cite{Souza:2015} for details) we get the balances 
\begin{align}
0 &= -\langle x'^2 \rangle_T + \rho'\langle x' y'\rangle_T + O(T^{-1}),\label{eq:xbalance}\\
0 &= \langle x'y' \rangle_T  - \langle y'^2\rangle_T - \beta\langle z'^2\rangle_T +  \frac{\varepsilon_y}{\rho'}\langle y' \rangle_T + O(T^{-1}),\label{eq:ybalance}\\
0 &= - \langle x'y'\rangle_T + \beta \langle z' \rangle_T + O(T^{-1}).\label{eq:zbalance}
\end{align}

We now write $z' = z_0 + \zeta(t)$ where $z_0 = \frac{\rho' - 1}{\rho'}$ is the so-called background component. Substituting into Equations (\ref{eq:ybalance}) and (\ref{eq:zbalance}) we get
\begin{align}
0 &= \langle x'y' \rangle_T - \langle y'^2\rangle_T -\beta z_0^2 - 2\beta z_0\langle\zeta\rangle_T - \beta\langle \zeta^2\rangle_T \\ & \qquad + \frac{\varepsilon_y}{\rho'}\langle y' \rangle_T + O(T^{-1}),\label{eq:ybackground}\\
0 &= - \langle x'y'\rangle_T + \beta z_0 + \beta \langle \zeta \rangle_T + O(T^{-1}).\label{eq:zbackground}
\end{align}

Taking (\ref{eq:ybackground}) + $2z_0\times$(\ref{eq:zbackground}) we find
\begin{equation}
\begin{aligned}
0 &= (1 - 2z_0)\langle x'y'\rangle_T -\langle y'^2\rangle_T - \beta\langle\zeta^2\rangle_T + \beta z_0^2 \\ & \qquad + \frac{\varepsilon_y}{\rho'}\langle y'\rangle_T + O(T^{-1}).
\end{aligned}
\end{equation}

Thus far our derivation is identical to that in \cite{Souza:2015}, but we now have the extra term $\varepsilon_y\langle y'\rangle_T$ which prevents us from completing the square as did the authors. Instead we eliminate the extra term by completing the square with respect to $x'$ and $y'$, and $y'$ and $\varepsilon_y$. First rewrite $\langle y'^2 \rangle_T = \delta \langle y'^2 \rangle_T + (1  - \delta)\langle y'^2\rangle_T$ where $\delta \in (0,1)$ so that
\begin{equation}
\begin{aligned}
0 & = 
\frac{2 - \rho'}{\rho'}\langle x'y'\rangle_T - \delta\langle y'^2\rangle_T - (1 - \delta)\langle y'^2\rangle_T - \beta\langle\zeta^2\rangle_T \\ & \qquad + \beta\frac{(\rho' - 1)^2}{\rho'^2} + \frac{\varepsilon_y}{\rho'}\langle y'\rangle_T  + O(T^{-1}),
\end{aligned}
\end{equation}
or multiplying through by $\rho'$ and rearranging,
\begin{equation}
\begin{aligned}
\rho'\langle x'y'\rangle_T &= 2\langle x'y'\rangle_T - \rho'\delta\langle y'^2\rangle_T - \rho'(1 - \delta)\langle y'^2\rangle_T \\ & \quad - \rho'\beta\langle\zeta^2\rangle_T  + \beta\frac{(\rho' - 1)^2}{\rho} + \varepsilon_y\langle y'\rangle_T + O(T^{-1}).
\end{aligned}
\end{equation}

Adding zeros in the form $\frac{1}{\rho\delta}$(\ref{eq:xbalance}) $ = {-\frac{1}{\rho\delta}\langle x'^2\rangle + \frac{1}{\delta}\langle x'y'\rangle}$, and  ${\frac{\varepsilon_y^2}{4\rho'(1 - \delta)} - \frac{\varepsilon_y^2}{4\rho'(1 - \delta)}}$ we find
\begin{equation}
\begin{aligned}
(\rho' - \frac{1}{\delta})\langle x'y'\rangle_T &\leq - \Big\langle (\sqrt{\rho'\delta}y' - \frac{1}{\sqrt{\rho'\delta}}x')^2\Big\rangle \\ & \quad - \Big\langle (\sqrt{\rho'(1 - \delta)}y' - \frac{\varepsilon_y}{2\sqrt{\rho'(1 - \delta)}})^2\Big\rangle \\ & \quad + \beta\frac{(\rho' - 1)^2}{\rho'} + \frac{\varepsilon_y^2}{4\rho'(1 - \delta)} + O(T^{-1}))
\\ & \leq
\beta\frac{(\rho' - 1)^2}{\rho'} + \frac{\varepsilon_y^2}{4\rho'(1 - \delta)} + O(T^{-1}),
\end{aligned}
\end{equation}
which in the infinite time limit and the original variables gives the bound
\begin{equation}
\begin{aligned}
\langle xy\rangle &\leq \beta\frac{(\rho' - 1)^2}{(\rho' - \frac{1}{\delta})} + \frac{\varepsilon_y^2}{4(\rho' - \frac{1}{\delta})(1 - \delta)}
\\ & = 
\beta\frac{\delta(\rho' - 1)^2}{\delta\rho' - 1} + \frac{\delta\varepsilon_y^2}{4(\delta\rho' - 1)(1 - \delta)}
\\ & := U(\delta),
\end{aligned}
\end{equation}
where we recall $\rho' = \rho - \frac{\varepsilon_z}{\beta}$. To make this optimal we minimize $U$. The derivative is given by
\begin{equation}
\begin{aligned}
U'(\delta) &= \beta(\rho' - 1)^2\frac{(\delta\rho' - 1) - \delta\rho'}{(\delta\rho' - 1)^2} \\ & \quad + \frac{\varepsilon_y^2}{4}\frac{(\delta\rho' - 1)(1 - \delta) - \delta(\rho' + 1 - 2\rho'\delta)}{(\delta\rho' - 1)^2(1 - \delta)^2}.
\end{aligned}
\end{equation}

Setting this equal to zero and simplifying we find
\begin{equation}
\beta(\rho' - 1)^2 = \frac{\varepsilon_y^2}{4}\frac{\delta^2\rho' - 1}{(1 - \delta)^2}, 
\end{equation}
yielding the following quadratic in $\delta$:
\begin{equation}
0 = [\beta(\rho' - 1)^2 - \rho'\frac{\varepsilon_y^2}{4}]\delta^2 - 2\beta(\rho - 1)^2\delta + \beta(\rho' - 1)^2 + \frac{\varepsilon_y^2}{4}, 
\end{equation}
which has solutions
\begin{equation}
\delta_\pm =
\frac{\beta(\rho' - 1)^2\pm \frac{\varepsilon_y}{2}\sqrt{\rho'\frac{\varepsilon_y^2}{4} + \beta(\rho' - 1)^3}}{\beta(\rho' - 1)^2 - \rho'\frac{\varepsilon_y^2}{4}}.
\end{equation}

We choose the negative square root so that $\delta\in(0,1)$ (one can easily check this is also the minimizer.) In the limit $\varepsilon_y,\varepsilon_z\rightarrow 0$ we find $\delta \rightarrow 1$ which recovers the bound $\langle xy\rangle \leq \beta(\rho - 1)$.

\bibliography{Lorenz_NoiseRefs}

\end{document}